\documentclass[preprint,3p,twocolumn]{elsarticle}

\usepackage[utf8]{inputenc}
\usepackage{graphicx}
\usepackage{tabularx}
\usepackage{amsmath}
\usepackage{amssymb}
\usepackage[version=4]{mhchem}
\usepackage{siunitx}
\biboptions{sort&compress}

\journal{XXX}
\begin{document}

\begin{frontmatter}

\title{Analyzing Iron Dust Bunsen Flames using Numerical Simulations}
\author[STFS]{T. Hazenberg\corref{cor1}}
\ead{hazenberg@stfs.tu-darmstadt.de}
\author[STFS]{D. Braig}
\author[KIT]{M. A. Fedoryk}
\author[STFS]{J. Mich}
\author[KIT]{F. P. Hagen}
\author[KIT]{S. R. Harth}
\author[KIT]{B. Stelzner}
\author[STFS]{A. Scholtissek}
\author[KIT]{D. Trimis}
\author[STFS]{C. Hasse}

\cortext[cor1]{Please address correspondence to T. Hazenberg}
\address[STFS]{Technical University of Darmstadt, Department of Mechanical Engineering, Simulation of reactive Thermo-Fluid Systems, Otto-Berndt-Straße 2, 64287 Darmstadt, Germany}
\address[KIT]{Engler-Bunte-Institute, Combustion Technology, Karlsruhe Institute of Technology, Engler-Bunte-Ring 7, 76131 Karlsruhe, Germany}


\date{\today}

\begin{abstract}
    This article presents numerical simulations of an iron dust Bunsen flame.
    The results are validated against experimental results.
    The burning velocity is extracted from the 3D simulation results, as in the experiments.
    The agreement of the burning velocity between the model and experiment is the best to date for iron dust flames. 
    A comparison is performed between 3D and 1D simulations to improve our understanding of how the 3D Bunsen flame deviates from an ideal 1D flame.
    This comparison reveals that the co-flow mixes with the post-flame zone, increasing the oxygen concentration in the reaction layer, which increases the burning velocity.
    Moreover, the analysis also reveals that stretch and curvature affect the burning velocity.
    These results are valuable for the future development of experimental setups aimed at measuring the burning velocity.
\end{abstract}

\end{frontmatter}

\section{Introduction}
Moving away from fossil fuels to renewable energy sources provides several challenges.
One of these is introducing intermittent renewable energy sources, i.e., solar and wind, to the electricity grid while maintaining stable operation.
Renewable chemical energy carriers then become a requirement for temporary energy storage.
These can account for load fluctuations or transport energy from regions with high to low renewable energy potential.
An interesting category of carbon-free energy carriers, that have received increased attention lately, is metals, e.g.,~\cite{2007_Mignard_reviewspongeiron,2015_Bergthorson_DirectCombustionRecyclable,2018_Bergthorson_RecyclableMetalFuels}. 
Metal powders can be used to operate a carbon-free energy cycle.
First, the metal powder is combusted, releasing energy, and the combustion product are metal oxides.
The metal oxides are captured and reduced with green hydrogen (energy storage), returning the metal oxide to metal and closing the cycle.
A promising candidate for constructing such a metal-based energy cycle is iron, which is the focus of this work.

Compared to other solid fuels, like coal and biomass, research on iron powders for energy storage and supply is still in its infancy.
A lot of pioneering work on iron (and other metal fuels for heat and power supply) has been performed at McGill University in Canada, e.g.,~\cite{2022_Goroshin_SomeFundamentalAspects}. 
Energy from iron powder can be rapidly oxidized with air in iron dust flames, releasing the stored energy as heat. 
However, the operation of iron dust flames is non-trivial since several fundamental differences exist between such (non-volatile) solid fuel flames, and conventional gas flames as pointed out by Goroshin et al. \cite{2022_Goroshin_SomeFundamentalAspects}.
Among others, the particle temperature can significantly exceed the gas temperature; fuel and oxidizer concentration relate non-trivially as the fuel does not diffuse, and the flame structure becomes distorted due to the particles` thermal inertia.
Consequently, the burning velocity and temperature respond differently to the fuel-oxygen equivalence ratio than gas flames~\cite{2021_Hazenberg_StructuresBurningVelocities}. 
In the past, iron dust flames have been established experimentally in various configurations, such as counterflow flames \cite{2019_McRae_StabilizedFlatIron}, methane/iron hybrid Bunsen flames~\cite{2015_Julien_Flamestructureparticle}, conventional Bunsen-type flames \cite{2023_Fedoryk_ExperimentalInvestigationLaminar}, spherical flames~\cite{1998_Sun_Structureflamespropagating,2000_Sun_CombustionBehaviorIron}, methane/iron hybrid flat flame burners~\cite{2024_Hulsbos_HeatFluxMethod}, lifted flames \cite{2024_Krenn_Evaluationnovelmeasurement}, and top-fired swirled tornado flames \cite{2023_Baigmohammadi_TowardsUtilizationIron}. 
Even though several stable flames could be operated in the lab, important flame characteristics, such as the flame stabilization mechanisms, are not thoroughly understood and warrant further research.

Parallel to the experimental work, significant progress has been made in the model development of metal dust flames.
In one of the earlier numerical works by Soo et al.~\cite{2017_Soo_ThermalStructureBurning}, they developed a single-particle model and used this to investigate the transient propagation of a hypothetical dust flame.
Later, Hazenberg and van Oijen~\cite{2021_Hazenberg_StructuresBurningVelocities} extended this model with temperature-dependent transport properties and based the particle model properties on that of iron; they than simulated the first 1D iron flames.
Several model variations and studies have been performed with this model as a base.
Van Gool et al.~\cite{2023_Gool_ParticleEquilibriumComposition} extended the model with temperature-dependent thermodynamic properties for iron and considered oxidation to the highest oxidation state (\ce{Fe2O3}).
The utilized thermodynamic table was later improved by Mich et al.~\cite{2024_Mich_Modelingoxidationiron} by accounting for non-ideal mixing enthalpy and entropy to capture the ``reactive cooling'' after peak particle temperature.
The original model and variants (inspired by~\cite{2022_Thijs_Improvementheatmass,2023_Gool_ParticleEquilibriumComposition,2022_Mi_quantitativeanalysisignition,1996_Goroshin_Quenchingdistancelaminar}) were used by Ravi et al.~\cite{2023_Ravi_Flamestructureburning,2023_Ravi_Effectparticlesize} and Mich et al.~\cite{2023_Mich_ComparisonMechanisticModels} to investigate the influence of particle size distributions.
The first to explore 3D effects on a 1D flame was van Gool et al.~\cite{2024_Gool_Numericaldeterminationiron}, who studied the impact of flame stretch using a 1D approximation to the counterflow setup.
Until now, only two numerical studies of laminar flame propagation have been performed in 3D: 1) Wen et al.~\cite{2023_Wen_Numericalmodelingpulverized} they studied the counterflow setup of McRae et al.~\cite{2019_McRae_StabilizedFlatIron}, and 2) Ramaekers et al.~\cite{2025_Ramaekers_influenceradiativeheat} they studied the impact of radiation on the tube flames of Tang et al.~\cite{2009_Tang_FlamePropagationQuenching,2011_Tang_Modesparticlecombustion}.
Rather differently, but not less useful, Vance et al.~\cite{2024_Vance_Flamepropagationmodes,2024_Vance_Flamepropagationmodesa,2024_Vance_numericalanalysismulti} have studied flame propagation using boundary layer resolved simulations, providing useful insights into the discrete regime~\cite{2009_Tang_Effectdiscretenessheterogeneous,2011_Goroshin_Reactiondiffusionfronts,2012_Tang_Propagationlimitsvelocity}.

In this article, we will present 3D numerical simulations of the experimental Bunsen flame of Fedoryk et al.~\cite{2023_Fedoryk_ExperimentalInvestigationLaminar}.
To this end, a point particle method in OpenFOAM\textregistered{} is utilized.
The particle model is that by Hazenberg and van Oijen~\cite{2021_Hazenberg_StructuresBurningVelocities}, with the thermodynamic properties and Stefan flow of Mich et al.~\cite{2023_Mich_ComparisonMechanisticModels}.
A particle size distribution is utilized in the experiments, which we approximate with a log-hyperbolic distribution function~\cite{1977_BarndorffNielsen_Exponentiallydecreasingdistributions}.
Our goal is twofold:
First and foremost, we would like to validate our numerical model against experiments, and second, we would like to identify if the Bunsen flame can, locally, be approximated as a 1D flame.
This article is split into several sections: first, we provide a short description of the experimental setup; second, the numerical model is introduced; third, the results are presented; and finally, we give an outlook and conclusions.

\section{Experimental Bunsen flame setup}
\begin{figure}
    \centering
    \includegraphics{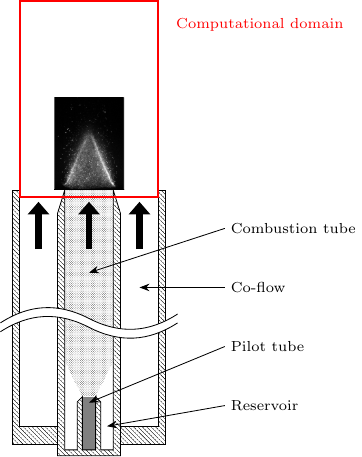}
    \caption{Sketch of the experimental configuration presented by Fedoryk et al.~\cite{2023_Fedoryk_ExperimentalInvestigationLaminar}. 
    Several of the relevant parts of the setup are indicated. 
    The computational domain, which extends \SI{8}{\centi\meter} above the combustion tube and co-flow outlet, is marked in red.
    A picture of the experimental flame, scaled correctly with respect to the computational domain, is inserted as a reference.}
    \label{fig:expSetup}
\end{figure}

The experimental setup utilizes iron powder of purity greater than 99.5\% (PMCtec GmbH, type YTF-HY2) as a fuel, which is contained in a cylindrical tube and shifted upwards by a piston connected to a stepper motor as shown in Fig.~\ref{fig:expSetup}. 
The dispersion of the iron powder occurs in a so-called air-knife seeder in a small gap of approximately \SI{30}{\micro\meter} height with high gas velocity. 
Mass flow controllers control the air mass flow in the seeder/burner and co-flow stream.

Fig.~\ref{fig:expSetup} further shows the burner setup, which consists of two concentric tubes: 
An inner pilot tube and an outer combustion tube with a diameter of \SI{20.5}{\milli\meter} and length of \SI{35}{\centi\meter} downstream from the pilot tube outlet. 
As the iron powder suspension exits the pilot tube, the flow slows down, causing some particles to fall.
It has also been observed that a few particles can also stick to the wall of the larger combustion tube before dropping. 
The falling powder is collected in a separate reservoir, minimizing fluctuations in the particle seeding. 
The outlet tube of the burner is surrounded by a co-flow with a larger diameter to enhance flame stability and protect the flame from external influences. 
Both the co-flow and combustion air were supplied at room temperature.
The mean inlet velocity of the co-flow and the combustion pipe are identical and can vary between \num{25} and \SI{45}{\centi\meter\per\second}.
A more detailed description can be found in \cite{2023_Fedoryk_ExperimentalInvestigationLaminar}.

\section{Numerical methods}
In the following, we present the framework for the numerical simulation of iron-air Bunsen flames.
The governing equations include both gas and particulate phase modeling. 
After the governing equations, the relevant boundary conditions are introduced. 
Thereafter, details about the computational setup are provided.

\subsection{Euler-Lagrange framework}
The iron-air Bunsen flame is simulated with an OpenFOAM\textregistered{}-based CFD code, utilizing the Euler-Lagrange framework. 
Iron microparticles are modeled as point particles according to the particle-source-in-cell approach. 
For the gas phase, the usual governing equations for mass, momentum, enthalpy, and chemical species read:
\begin{equation}
    \frac{\partial}{\partial t} \rho_{\mathrm{g}} 
  + \frac{\partial}{\partial x_i}\left(\rho u_{\mathrm{g},i}\right) 
  = 
    S_{\mathrm{prt}, m} \mathrm{,}
\end{equation}
\begin{align}
    \frac{\partial}{\partial t}\left(\rho_{\mathrm{g}} u_{\mathrm{g},i}\right) 
  &+ \frac{\partial}{\partial x_j}\left(\rho_{\mathrm{g}} u_{\mathrm{g,i}} u_{\mathrm{g},j}\right) \\
  &=
  - \frac{\partial}{\partial x_i}p 
  + \frac{\partial}{\partial x_j}\tau_{i j} 
  + \rho_{\mathrm{g}} g_i+S_{\mathrm{prt}, u_{\mathrm{g},i}} \nonumber \mathrm{,}
\end{align}
\begin{align}
    \frac{\partial}{\partial t}(\rho_{\mathrm{g}} h)
  & + \frac{\partial}{\partial x_j}\left(\rho_{\mathrm{g}} u_{\mathrm{g},j} h\right) \\
  & = 
    \rho_{\mathrm{g}} u_{\mathrm{g},j} g_j
  - \frac{\partial }{\partial x_j}q_j
  + S_{\mathrm{prt}, h} \nonumber
\end{align}
and
\begin{align}
    \frac{\partial}{\partial t}\left(\rho_{\mathrm{g}} Y_{\mathrm{g},k}\right)
  & + \frac{\partial}{\partial x_j}\left(\rho_{\mathrm{g}} u_{\mathrm{g},j} Y_{\mathrm{g},k}\right) \\
  & = 
    \frac{\partial}{\partial x_j}\left(\rho_{\mathrm{g}} D_k \frac{\partial }{\partial x_j}Y_{\mathrm{g},k}\right) 
  + S_{\mathrm{prt}, Y_{\mathrm{g},k}} \mathrm{,} \nonumber
\end{align}
where $\rho_{\mathrm{g}}$ is the density, 
$u_{\mathrm{g},i}$ is the flow velocity in spatial dimension $x_i$, 
$p$ is the pressure, $\tau_{ij}$ is the stress tensor, 
$g_i$ is the gravity, $h$ is the enthalpy, $q_i$ is the heat flux, 
$Y_{\mathrm{g},k}$ is the mass fraction and $D_k$ is the diffusivity of species k. 
Finally, in the above equations, $S_{\mathrm{prt}}$ represents exchange terms between the continuous gas and the dispersed solid phase, which are defined later. 

\subsection{Particle Model}
We use a particle model based on the work of Soo et al.~\cite{2015_Soo_ReactionParticleSuspension,2017_Soo_ThermalStructureBurning}, which is adapted to iron powder by Hazenberg and van Oijen~\cite{2021_Hazenberg_StructuresBurningVelocities} and later applied to one-dimensional polydisperse iron-air flames by Mich et al.~\cite{2023_Mich_ComparisonMechanisticModels} and Ravi et al.~\cite{2023_Ravi_Effectparticlesize}. 
Mich et al.~\cite{2023_Mich_ComparisonMechanisticModels} extended the particle model with thermodynamic properties that depend on temperature, which is the framework utilized in this article.
In this model, the evolution of the particle mass is given by
\begin{equation}
  \frac{\mathrm{d} m_{\mathrm{p}}}{\mathrm{d} t} = \dot{m}_{\ce{ox}}
  \label{eq:mass_evolution}
\end{equation}
and that of the iron mass by
\begin{equation}
  \frac{\mathrm{d} m_{\mathrm{p},\ce{Fe}}}{\mathrm{d} t} = -\frac{1}{s} \dot{m}_{\ce{ox}} \mathrm{,}
\end{equation}
where $\dot{m}_{\ce{ox}}$ is the oxidation rate of the particle and $s$ is the stoichiometric ratio of iron to oxygen.
$s$ is given by $M_{\ce{Fe}}/M_{\ce{O2}}=55.85/15.99$ when complete oxidation up to $\ce{FeO}$ is assumed and higher oxides are neglected~\cite{2021_Hazenberg_StructuresBurningVelocities}.
The particle oxidation rate can be controlled by surface kinetics,
\begin{equation}
  \dot{m}_{\ce{O2},\mathrm{kin}} = \rho_{\mathrm{s}} Y_{\mathrm{s},\ce{O2}} A_{\mathrm{p}} k_{\infty} \exp \left(\frac{-T_{\mathrm{a}}}{T_{\mathrm{p}}}\right) \mathrm{,}
  \label{Eq:O2_kinetic_rate}
\end{equation} 
or the diffusion of oxygen through the boundary layer (including Stefan flow),
\begin{equation}
  \dot{m}_{\ce{O2},\mathrm{diff}} = -A_{\mathrm{p}} \frac{Sh (\rho D_{\ce{O2}})_{\mathrm{f}} }{ d_{\mathrm{p}} }\mathrm{ln}(1+B_{\mathrm{m}}) \mathrm{.}
  \label{Eq:O2_diffusion_rate}
\end{equation}
In the above equations, $Y_{\ce{O2}}$ is the mass fraction of oxygen, $A_{\mathrm{p}}$ is the particle surface area, $d_{\mathrm{p}}$ is the particle diameter, $k_{\infty}$ is the pre-exponential factor of the surface reaction rate, $T_{\mathrm{a}}$ is the surface reaction activation temperature, $Sh$ is the Sherwood number, $(\rho D_{\ce{O2}})_{\mathrm{f}}$ is the diffusivity of oxygen inside the film layer and $B_{\mathrm{m}}$ is the Spalding number given by:
\begin{equation}
  B_{\mathrm{m}} = \frac{Y_{\mathrm{s},\ce{O2}} - Y_{\mathrm{g},\ce{O2}}}{1 - Y_{\mathrm{s},\ce{O2}}} \mathrm{.}
\end{equation}
Subscripts $_{\mathrm{s}}$, $_\mathrm{f}$ and $_{\mathrm{g}}$ indicate evaluated at the surface, inside the boundary layer and in the bulk gas phase.
The overall oxidation rate of a particle is computed by assuming a steady-state boundary layer, such that
\begin{equation}
  \dot{m}_{\mathrm{kin}} = \dot{m}_{\mathrm{diff}} \mathrm{,}
\end{equation}
which introduces an algebraic relation for $Y_{\mathrm{s},\ce{O2}}$.
In our numerical solver, we approximated Eq.~\ref{Eq:O2_diffusion_rate} by a second-order polynomial function, such that no iterative scheme is required to solve for $Y_{\mathrm{s},\ce{O2}}$.
The mass, volume, and surface area of a particle can be related by assuming that the particle is spherical and the particle density ($\rho_{\mathrm{p}}$) follows the following simple mixing rule
\begin{equation}
  \frac{1}{\rho_{\mathrm{p}}} = \frac{Y_{\mathrm{p},\ce{Fe}}}{\rho_{\ce{Fe}}} + \frac{Y_{\mathrm{p},\ce{FeO}}}{\rho_{\ce{FeO}}} \mathrm{,}
\end{equation}
where $Y_{\mathrm{p},i}$ are the species mass fractions inside the particle, and $\rho_{i}$ are the temperature-dependent densities of \ce{Fe} and \ce{FeO}.

The evolution of the particle temperature is described by a temperature equation,
\begin{equation}
  m_{\mathrm{p}} c_{\mathrm{p},\mathrm{p}} \frac{\mathrm{d} T_{\mathrm{p}} }{\mathrm{d} t} = \dot{q}_{\mathrm{conv}} + \dot{q}_{\mathrm{rad}} + \dot{q}_{\mathrm{reac}} + \dot{m}_{\ce{O2}} h_{\mathrm{s},\ce{O2}},
  \label{eq:enthalpy_evolution}
\end{equation}
where $c_{\mathrm{p},\mathrm{p}}$ is particle heat capacity, including the latent heat of phase changes, and the right-hand side includes convective heat transport, radiative heat loss, heat of reaction and energy transfer due to enthalpy transport.
In the appendix we have added the derivation of the temperature equation from the enthalpy equation presented in earlier works, e.g.,~\cite{2021_Hazenberg_StructuresBurningVelocities,2023_Mich_ComparisonMechanisticModels,2023_Ravi_Effectparticlesize,2023_Ravi_Flamestructureburning,2023_Gool_ParticleEquilibriumComposition,2023_Gool_ParticleEquilibriumComposition}.
The convective heat transfer includes a correction for the Stefan flow and is provided by
\begin{equation}
  \dot{q}_{\mathrm{conv}} = - A_{\mathrm{p}} \frac{Nu \lambda_{\mathrm{f}}}{d_{\mathrm{p}}}\left(T_{\mathrm{p}}-T_{\mathrm{g}}\right) \frac{\ln \left(1+B_{\mathrm{t}}\right)}{B_{\mathrm{t}}},
\end{equation}
the radiative heat loss is given by
\begin{equation}
  \dot{q}_{\mathrm{rad}} = -\epsilon_{\mathrm{p}} \sigma A_{\mathrm{p}} \left(T_{\mathrm{p}}^4-T_{\mathrm{g}}^4\right)\mathrm{,}
\end{equation}
and the heat release of reaction by
\begin{equation}
  \dot{q}_{\mathrm{reac}} = - h_{\ce{Fe}} \frac{\mathrm{d} m_{\mathrm{p},\ce{Fe}}}{\mathrm{d} t} - h_{\ce{FeO}} \frac{\mathrm{d} m_{\mathrm{p},\ce{FeO}}}{\mathrm{d} t} \mathrm{,}
\end{equation}
where $B_{\mathrm{t}}$ is the energy Spalding number, $Nu$ is the Nusselt number, $\epsilon_{\mathrm{p}}$ is the emissivity of the particle and $\sigma$ is the Stefan-Boltzmann constant.
In the boundary layer, it is assumed that the Lewis number is unity, i.e., $B_{\mathrm{t}}=B_{\mathrm{m}}$, and the flow is stationary, i.e., $Sh=Nu=2$. 
Finally, the particle thermodynamics (phase densities, heat capacities, etc.) are described with correlations from the NIST database~\cite{1998_Chase_NISTJANAFThemochemical}, including transitions between phases.

The OpenFOAM\textregistered{} libraries handle particle tracking; the evolution of the particle velocity is described by
\begin{equation}
  m_{\mathrm{p}} \frac{\mathrm{d} u_{\mathrm{p},i}}{\mathrm{d} t} = m_{\mathrm{p}} g_i + F_{\mathrm{d}, i} + \left(u_{\mathrm{g},i} - u_{\mathrm{p},i}\right) \frac{\mathrm{d} m_{\mathrm{p}}}{\mathrm{d} t} \mathrm{,}
\label{eq:velocity_evolution}
\end{equation}
the last term is due to the change in particle momentum as a result of the mass change, and $F_{\mathrm{d},i}$ is the drag force provided by
\begin{equation}
  F_{\mathrm{d}, i}=\frac{1}{2} \rho_{\mathrm{g}} \left|u_{\mathrm{g}, i}-u_{\mathrm{p}, i}\right| \left(u_{\mathrm{g}, i}-u_{\mathrm{p}, i}\right) C_{\mathrm{d}} A_{\mathrm{p},\mathrm{c}} \mathrm{,}
\end{equation}
with
\begin{equation}
  C_{\mathrm{d}}=\frac{24}{Re_{\mathrm{p}, i}}\left(1+\frac{1}{6} Re_{\mathrm{p}, i}^{2 / 3}\right) \mathrm{,}
\end{equation}
where $A_{\mathrm{p},\mathrm{c}}$ the cross-sectional area and $Re_{\mathrm{p}}$ the particle Reynolds number. 

Based on Eqs.~\ref{eq:mass_evolution},~\ref{eq:velocity_evolution}~and~\ref{eq:enthalpy_evolution}, the exchange terms for a numerical cell with the gas phase are defined as
\begin{equation}
  S_{\mathrm{prt}, m} = -\frac{1}{V_{\mathrm{cell}}} \sum_m \left(\frac{ \mathrm{d} m_{\mathrm{p}}}{ \mathrm{d} t}\right)_{j} n_{j} \mathrm{,}
\end{equation}
\begin{equation}
  S_{\mathrm{prt}, u_i} = -\frac{1}{V_{\mathrm{cell}}} \sum_m \left(F_{\mathrm{d},i}\right)_j n_j \mathrm{,}
\end{equation}
\begin{equation}
  S_{\mathrm{prt}, h} = -\frac{1}{V_{\mathrm{cell}}} \sum_m \left( \dot{q}_{\mathrm{conv}} + h_{\ce{O2}, \mathrm{s}}\frac{\mathrm{d} m_{\mathrm{p}}}{\mathrm{d} t}\right)_{j} n_{j} \mathrm{,}
\end{equation}
\begin{equation}
  S_{\mathrm{prt}, Y_{\ce{O2}}} = -\frac{1}{V_{\mathrm{cell}}} \sum_m \left(\frac{\mathrm{d} m_{\mathrm{p}}}{\mathrm{d} t}\right)_j n_j
\end{equation}
and
\begin{equation}
  S_{\mathrm{prt},i\neq\ce{O2}} = 0
\end{equation}
where $V_{\mathrm{cell}}$ is the volume of the corresponding cell, the subscript $m$ represents the parcels inside a cell, subscript $j$ is the index of the $j$'th parcel and $n_j$ is the number of particles per parcel for the $j$'th parcel. 

\subsection{Numerical Setup}
The three-dimensional computational domain for the Bunsen flame is indicated in Fig.~\ref{fig:expSetup}. 
It is discretized with \num{1.1} million cells with inlets for the main pipe flow (particles + air) and co-flow (air) at the bottom of the domain.
The outlets are located at the sides and top of the domain; a small part of the inlet pipe is included in the model to which no-slip wall boundary conditions are applied. 
An overview of the boundary conditions is provided in Table~\ref{tab:BC}. 

\begin{table*}
  \centering
  \caption{Boundary conditions specified for the computational domain.}
  \begin{tabular}{llll}
    \hline
    Surface & T & U & p \\ \hline
    Fuel & fixed value & fixed value  & fixed value  \\ 
    Co-flow	  & fixed value & fixed value & zero gradient  \\ 
    Wall  & fixed value & no slip & zero gradient  \\ 
    Surroundings  & zero gradient & inlet/outlet & wave transmissive  \\ 
  \end{tabular}
  \label{tab:BC}
\end{table*}

\begin{figure}
  \centering
  \includegraphics{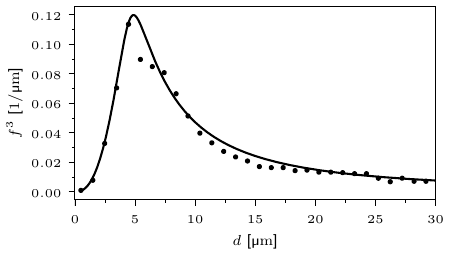}
  \caption{The experimental particle size distribution and the fitted particle size distribution. 
  $f^3$ indicates the mass-based particle size probability density function.}
  \label{Fig:PSD}
\end{figure}

The particle size distribution of the powder utilized in the experiments of Fedoryk et al.~\cite{2023_Fedoryk_ExperimentalInvestigationLaminar} is measured both in-situ and with a Camsizer.
We believe that the particle size distribution characterized by a Camsizer is more reliable than the in-situ measurements in the article, as it does not depend on the flow conditions of the gas inside the Bunsen setup.
Using the Camsizer, three measurements are made of the particle size distribution, each with a different definition of the particle size.
The average of the three distributions is taken to which a log-hyperbolic~\cite{1977_BarndorffNielsen_Exponentiallydecreasingdistributions} distribution is fitted with satisfactory accuracy ($\mathrm{adj.}\:R^2 = 0.99$), see Fig.~\ref{Fig:PSD}.
Only particles between \num{0.5} and \SI{30}{\micro\meter} are included in the simulation; the distribution is renormalized accordingly.
In the appendix, we provide the particle size distributions, more details on the particle size definitions, and details on the utilized particle size distribution.

\begin{figure}
  \centering
  \includegraphics{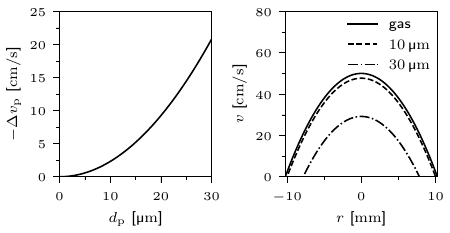}
  \caption{Left: the terminal slip velocity of the particles as a function of their initial diameter. 
  Right: The inlet velocity profile for the gas and various particle sizes.
  Particles with negative velocity at the inlet are removed from the simulation.}
  \label{fig:Inlet_velocity}
\end{figure}

The inlet at the central pipe is modeled with a fully developed pipe flow close to the experimentally observed inlet velocity profile~\cite{2023_Fedoryk_ExperimentalInvestigationLaminar}.
Accordingly, at the inlet, the particles are injected at the terminal velocity, i.e., $F_{\mathrm{d},i} = m_{\mathrm{p}} g_i$, and particles whose velocity is negative at the inlet are removed, see Fig.~\ref{fig:Inlet_velocity} for reference.
We assume that the concentration of the particles above the inlet is uniform, such that the mass flux of particles varies as a function of radius from the inlet center together with the inlet velocity.
The local mass flux of the particles is scaled such that the target particle loading is reached if the particle and gas velocity are equal.

Finally, parcels are used instead of particles to reduce the computational load.
The smallest parcels (\SI{0.5}{\micro\meter}) contain around \num{2000} particles, and any parcel larger than \SI{6}{\micro\meter} is modeled as a single particle.
In between \num{0.5} and \SI{6}{\micro\meter}, the number of particles per parcel is chosen such that the total initial mass of the particles inside a particle is equal to that of a \SI{6}{\micro\meter} particle.
As a result of these settings, the domain contains over \num{3}~million parcels when the flame has reached a (quasi-)steady state.
These parcel settings provide a good compromise between accuracy and computational costs.

\section{Results}
The numerical simulations presented in this article are performed at four of the conditions presented in Fedoryk et al.~\cite{2023_Fedoryk_ExperimentalInvestigationLaminar}.
Two different equivalence ratios are presented based on \ce{Fe2O3} as final product, $\phi_{\ce{Fe2O3}} = \num{1.0}$~and~\num{1.5}, and two different average inlet velocities $\bar{v_{\mathrm{in}}}=\num{25}$~and~\SI{40}{\centi\meter\per\second}.
A steady-state flame was obtained and then sampled at \SI{100}{\hertz} for half a second for all the simulations conducted.
These time-dependent results show that the flame remains highly stable and cylindrical symmetric; as such, we will not show time-averaged or time-depended behavior, as this did not modify the results.
In this section, we will first visually compare the simulation results against the experiments; second, we will analyze the numerical burning velocity like the experimental flame; third, the flame structure will be compared against 1D numerical results.
Finally, the burning velocities of the 3D simulations, 1D simulations and experiments are compared.

\subsection{Analysis of the experiments}
\begin{figure}
  \centering
  \includegraphics{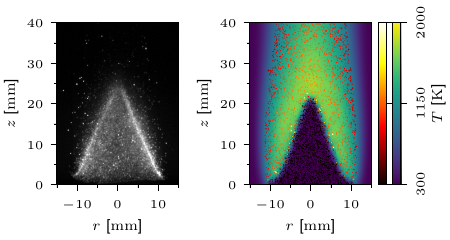}
  \caption{Comparison between the experimental iron Bunsen flame and the numerical result, at $\bar{v}_{\mathrm{in}}=\SI{25}{\centi\meter\per\second}$ and $\phi_{\ce{Fe2O3}}=1.0$. 
  Left: An image of an instantaneous snapshot of the flame is shown. 
  Right: A slice of the numerical domain showing gas and particle temperatures. 
  The color bar for the particles has been clipped to 2000 K to provide reasonable color resolution in the post-flame zone.}
  \label{Fig:Comparison_experiment}
\end{figure}

In Fig.~\ref{Fig:Comparison_experiment}, a visual comparison is made between the experimental and numerical result at $\phi_{\ce{Fe2O3}}=1.0$ and $\bar{v}=\SI{25}{\centi\meter\per\second}$.
From the experiment, an instantaneous snapshot is shown, while the gas-phase and particle temperature are shown for the numerical results.
The snapshot of the experiments presented here has slightly different dimensions than Fig.~4 presented in Fedoryk et al.~\cite{2023_Fedoryk_ExperimentalInvestigationLaminar} due to a scaling error in Fig.~4 of~\cite{2023_Fedoryk_ExperimentalInvestigationLaminar}.
This scaling error only affects the z-axis representation in the figure and does not impact the results, as they are based on the original data.
The overall shape of the flame in the numerical result is comparable to that of the experimental result.
At first sight, the experimental flame appears slightly larger ($\pm \SI{25}{\milli\meter}$) than the numerical flame ($\pm \SI{22.5}{\milli\meter}$).
However, we should be careful drawing conclusions from these images as two related but different quantities are compared.
In the experiment, a line-of-sight measurement is taken from the radiating iron particles.
In the numerical result, a contour of the gas temperature and a scatter of the particles sized by their diameter and colored by temperature is shown.
It is hard to make a direct comparison, as the experiment shows light intensity from the radiating particles, and the numerical result shows the contours of the gas temperature.
Thus, further analysis is required to compare the experimental and numerical results reasonably.

To better compare the numerical result to the experimental results, the laminar burning velocity is estimated from the experiments.
For this, three different methods are applied:
1) The area of an iso-contour of temperature $A^{T^*}_{\mathrm{ref}}$ is compared with the area of the fuel inlet ($A_{\mathrm{in}}$).
The global laminar burning velocity can then be obtained from
\begin{equation}
  S_{\mathrm{L},\mathrm{A}} = \bar{v_{\mathrm{in}}} \frac{A_{\mathrm{in}}}{A^{T^*}_{\mathrm{ref}}} \mathrm{.}
  \label{Eq:Flame_speed_area_based}
\end{equation}
2) The flame front is reconstructed from a temperature iso-contour at $T^*$.
The reconstructed flame front is utilized to obtain the local flame angle ($\theta(r)$).
This local flame angle can be compared against the inlet velocity ($v_{\mathrm{in}}(r)$) at the same radius.
The local flame speed can then be obtained from
\begin{equation}
  S_{\mathrm{L},\theta} = v_{\mathrm{in}}(r) \cos \left(\theta(r)\right) \mathrm{.}
  \label{Eq:Flame_speed_angle_based}
\end{equation}
This method was also utilized in the experiments.
3) The same reconstructed flame front as in 2) is utilized to construct a local flame-attached coordinate system.
The local velocity vector is then transformed onto this flame-attached coordinate system to obtain the flame-normal ($v_{\eta}$) and flame-tangential ($v_{\xi}$) velocity.
To obtain the laminar burning velocity, the flame-normal velocity is corrected for gas expansion via a temperature correction, i.e.,
\begin{equation}
  S_{\mathrm{L},\eta} = v_{\eta} \frac{T^*}{T_{\mathrm{in}}} \mathrm{.}
  \label{Eq:Flame_speed_flow_based}
\end{equation}
For this third method, care should be taken that the iso-contour of temperature is sufficiently low, such that no oxygen has been transferred from the gas phase to the particles.
From methods 2 and 3, an overall laminar burning velocity is obtained by taking the average between $r/R_{\mathrm{max}}=0.33$ and \num{0.66}.
In all three methods, we assume that the flame is cylindrical symmetric, such that circumferential averaging can be performed.
We will refer to method 1) as the area based method, to method 2) as the angle method and to method 3) as the flow method.

\begin{figure}
  \centering
  \includegraphics{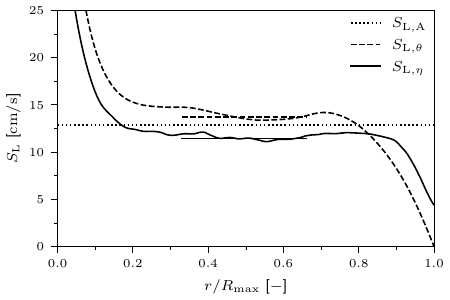}
  \caption{The laminar burning velocity is estimated from the numerical results of the Bunsen flame using three methods (Eq.~\ref{Eq:Flame_speed_area_based} to Eq.~\ref{Eq:Flame_speed_flow_based}).
  The area-based method only provides an overall burning velocity and is thus represented by a constant line.
  For both the angle and flow method, two lines are shown: a non-constant line, showing the local value, and a constant line between $r/R_{\max} = 0.33$ and $0.66$, showing the overall burning velocity.}
  \label{Fig:Flame_speed_1_25}
\end{figure}
The burning velocity is computed using the above three methods for the $\phi=1.0$ and $\bar{v}=\SI{25}{\centi\meter\per\second}$ case.
Results from this computation, utilizing the $T^*=\SI{400}{\kelvin}$ iso-contour, are depicted in Fig~\ref{Fig:Flame_speed_1_25}.
The graph shows the local burning velocity using the angle and flow method and the global burning velocity of all three methods as a constant line.
For the locally computed burning velocity, the trend is similar to what is observed in the experiments (Fig. 6 of Fedoryk et al.~\cite{2023_Fedoryk_ExperimentalInvestigationLaminar}):
The maximum burning velocity in the center of the Bunsen flame is higher due to the strong curvature of the flame tip, and near the burner rim it tends to zero due to heat loss to the rim.
Between $r/R_{\max} = 0.33$ and \num{0.66}, the obtained burning velocity varies by less than \SI{1.5}{\centi\meter\per\second} for both methods.
Finally, the burning velocity obtained with all methods is slightly lower than the experimentally observed burning velocity (\num{12.5} to \SI{15.2}{\centi\meter\per\second}).
This comparison of the burning velocity shows that care must be taken with a simple visual comparison, as in Fig.~\ref{Fig:Comparison_experiment}, from which one would conclude that the burning velocity in the numerical simulation is higher than in the experiment.

\begin{figure}
  \centering
  \includegraphics{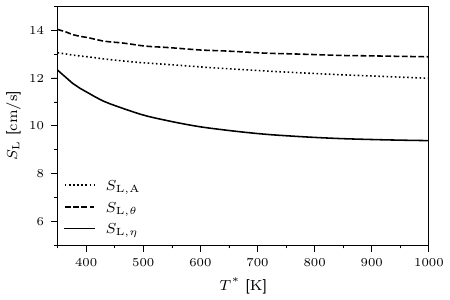}
  \caption{Dependency of the global laminar burning velocity on the chosen temperature iso-contour.
  See Fig.~\ref{Fig:Flame_speed_1_25} for local laminar burning velocity dependency on position.}
  \label{Fig:Var_Iso_Flame_speed_1_25}
\end{figure}
In Fig.~\ref{Fig:Flame_speed_1_25}, the utilized iso-contour of temperature is an arbitrary choice.
As Fedoryk et al. used the optical emission of the radiating particle, a higher iso-contour of temperature is likely much more representative.
As such, we have analyzed the impact of the utilized temperature iso-contour on the obtained global burning velocity.
In Fig~\ref{Fig:Var_Iso_Flame_speed_1_25}, the iso-contour of temperature varies from \num{350} to \SI{1000}{\kelvin}, and the global burning velocity using all three methods is computed for each.
The graph shows that the obtained burning velocity decreases for increasing temperature iso-contour.
In the case of the area and angle method, this decrease is relatively small, around \SI{1}{\centi\meter\per\second} over the entire range.
However, the local flow method shows a more significant decrease of around \SI{3}{\centi\meter\per\second}.
This decrease can be readily explained for the area and angle methods:
The temperature iso-contour becomes slightly larger/grows radially outward for increasing temperature. 
As such, the area increases for the area-based method, while for the angle-based method $v_{\mathrm{in}}(r)$ decreases for larger radial coordinates. 

For the flow-based method, several effects are at play:
As the inlet velocity profile is quadratic (fully developed pipe flow), the flame exhibits both areas of negative stretch for small $r/R_{\max}$ and positive stretch for large $r/R_{\max}$, see e.g.,~\cite{2014_Fu_Effectpreferentialdiffusion}.
In regions of positive stretch, one would expect that the flame speed obtained by Eq.~\ref{Eq:Flame_speed_flow_based} would decrease, as there is a net mass loss in the local flamelets, while in regions of negative stretch, the reverse would be true.
At the same time, as the iso-contour of temperature is increased, the radius of this contour also increases.
As a result, after correcting for temperature the flow velocity, must decrease within a flamelet to adhere to conservation, which is not accounted for in Eq.~\ref{Eq:Flame_speed_flow_based}.
Besides, curvature and stretch can also directly impact the mass burning rate.
The impact of either on the mass burning rate has not been studied extensively. 
The influence of stretch on the burning velocity has been investigated by van Gool et al.~\cite{2024_Gool_Numericaldeterminationiron}; they found that the burning velocity is increased by positive stretch.
Further analysis on each of these effects is not performed, as understanding how to extract best a stretch and curvature-free burning velocity from a Bunsen flame is outside the scope of this article.

The global burning velocity of the three other cases simulated will be compared later against the measured burning velocity and 1D results.
Based on this single comparison, we conclude that the numerical results show good agreement with the experimental results, or at least they do not show the typical factor of two lower compared to earlier comparisons~\cite{2021_Hazenberg_StructuresBurningVelocities,2025_Ramaekers_influenceradiativeheat}.
This agreement provides some validation of the models that have been developed for iron dust flames in recent years.

\subsection{Structures of 1D flames}
\begin{figure}
  \centering
  \includegraphics{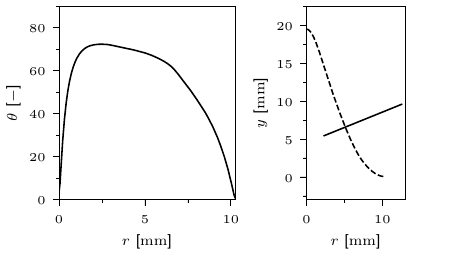}
  \caption{In the right frame, the reconstructed flame front is shown (based on the \SI{400}{\kelvin} iso-contour). 
  A local flamelet is drawn, at $r/R_{\max}=0.5$, orthogonal to the temperature iso-contour.
  In the left frame, the local angle of the reconstructed flame with respect to the upward direction is shown.}
  \label{Fig:Reconstructed_flame_front}
\end{figure}

For future comparisons, it is important to understand how well results from the experimental Bunsen flame can be compared to idealized 1D flames.
To investigate how well the experimental Bunsen flame compares to a 1D flame, a 1D flamelet is extracted from the 3D Bunsen flame.
This extracted flame is then compared against a 1D flame in similar conditions.
The flamelet is extracted from the Bunsen flame by reconstructing the flame front, as shown in the right frame of Fig.~\ref{Fig:Reconstructed_flame_front}.
Identical to the analysis of the previous section, the flamefront is reconstructed via the iso-contour of temperature.
Given this iso-contour, a local flame coordinate system can be defined at any position along the flame front.
The outward pointing direction, from here referred to as $\eta$, of this coordinate frame is considered to be a suitable flamelet.
In this manner, we have extracted a line at the position $r/R_{\max}=0.5$, the solid line in the right frame of Fig~\ref{Fig:Reconstructed_flame_front}.

Presenting the particle data on a line graph as a function of $\eta$ provides a challenge, as the particles are discrete.
A Gaussian kernel is utilized to convert the particle point data to a line graph. 
This Gaussian kernel averages (and smooths) the particle data in a small region around a position among the flamelet.
The utilized Gaussian kernel has a standard deviation of $\sigma=0.5\sqrt{2} \si{\milli\meter}$, and points further away than \SI{1}{\milli\meter} are excluded from the stencil.
As particles of various sizes behave very differently, the Gaussian kernel is applied to three bins of particles: $b^1$ ranges from \num{0.5} to \SI{2.5}{\micro\meter}, $b^2$ ranges from \num{10} to \SI{15}{\micro\meter}, and $b^3$ ranges from \num{20} to \SI{30}{\micro\meter}.
The gas phase is also smoothed with the same Gaussian kernel to ensure the relative widths of features in the gas and particles phase are conserved.

To conduct a 1D simulation at a representative condition, the gravity is reduced to account for the angle of the flame front.
The angle of the flame front as a function of $r/R_{\max}$ is shown in the left frame of Fig.~\ref{Fig:Reconstructed_flame_front}.
Note that this is the same angle used before in the angle method.
The flanks of the Bunsen flame are relatively straight, so the angle is nearly constant (roughly \num{70} degrees).
Based on this, we conduct 1D simulations with a gravity of $g = -9.81 \cos(70)$ \si{\meter\per\second\squared}.
The data of the 1D flames is then post-processed identically to the 3D flames to provide a fair comparison. 

\begin{figure}
  \centering
  \includegraphics{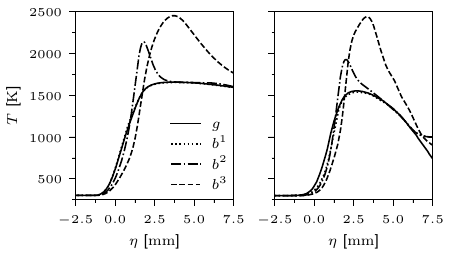}
  \caption{In the left frame, the temperature are obtained from a 1D flame simulation.
  In the right frame, the temperature profile among a flamelet retrieved from the 3D simulation; see Fig.~\ref{Fig:Reconstructed_flame_front} for the position. 
  The particle temperature is post-processed by binning them: $b^1$ ranges from \num{0.5} to \SI{2.5}{\micro\meter}, $b^2$ ranges from \num{10} to \SI{15}{\micro\meter}, and $b^3$ ranges from \num{20} to \SI{30}{\micro\meter}, and the solid lines gives the gas temperature.
  Smoothing using a Gaussian kernel is applied to both gas and particle temperature profiles for both the 1D and 3D results.}
  \label{Fig:Temperature_flamelet}
\end{figure}

In Fig.~\ref{Fig:Temperature_flamelet}, the temperature profiles of a 3D flame are compared against the 1D flame.
The smallest bin of particles appears to follow the gas phase closely and not significantly exceed it.
Unfortunately, this is an unavoidable side effect of the post-processed. 
The utilized Gaussian kernel removes features that are much smaller than the stencil size. 
Small particles have very short characteristic time scales, and thus, small features in a spatial sense; the kernel removes these features.
The result is that the graphed particle temperature closely follows the gas phase temperature.
Moreover, the binning of the particles also removes some features.

Several observations can be made when comparing the 1D and the 3D results.
First, the width of the flame front is relatively similar; both fronts start at around $\eta = \SI{0}{\milli\meter}$ and end at roughly $\eta = \SI{2.5}{\milli\meter}$.
Second, the maximum gas temperature achieved in both flames is roughly \SI{1500}{\kelvin}; the 1D simulation provides a slightly higher temperature, and the 3D simulation slightly lower.
Third, the position at which bins \num{2} and \num{3} achieve the maximum and the maximum temperatures are very similar for 1D and 3D.
Finally, in the post-flame zone $\eta > \SI{2.5}{\milli\meter}$, the temperature of the gas phase and the particles reduce much more quickly in the 3D simulation.
From observations one to three, we conclude that the flame front of the 3D flame behaves relatively well as a 1D, as the differences for $\eta <\SI{2.5}{\milli\meter}$ are minor.
However, the post-flame region is not accurately described by the 1D model, as the gas temperature reduces much quicker in the 3D result.
If we return to Fig.~\ref{Fig:Temperature_flamelet} and look at the solid line representing the extracted flamelet, it can be observed that this line extends beyond $r=R_{\max}=\SI{10.25}{\milli\meter}$.
This position lies in the cold gas of the co-flow, see Fig.~\ref{Fig:Comparison_experiment}, meaning that the co-flow is rapidly cooling the post-flame zone, explaining the differences between the 1D and 3D result.

\begin{figure}
  \centering
  \includegraphics{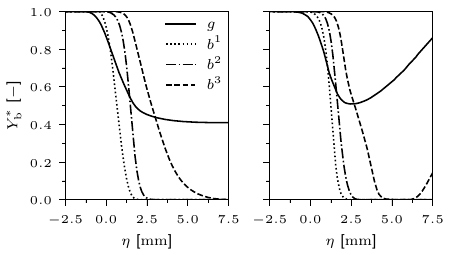}
  \caption{In the left frame, the unburned fraction profile is obtained from a 1D flame simulation.
  In the right frame, the unburned fraction profile among a flamelet retrieved from the 3D simulation; see Fig.~\ref{Fig:Reconstructed_flame_front} for the position.
  The unburned fraction of the gas phase is the normalized oxygen mass fraction, which is presented by the solid line.
  The particle unburned fraction ($Y_{\ce{Fe}}$), is shown by binning them, $b^1$ ranges from \num{0.5} to \SI{2.5}{\micro\meter}, $b^2$ ranges from \num{10} to \SI{15}{\micro\meter}, and $b^3$ ranges from \num{20} to \SI{30}{\micro\meter}.
  Smoothing using a Gaussian kernel is applied to both gas and particle unburned fraction profiles for both the 1D and 3D results.}
  \label{Fig:Unburned_fraction_flamelet}
\end{figure}

Similar to the temperature profile, also the unburned fraction profiles are graphed, see Fig.~\ref{Fig:Unburned_fraction_flamelet}.
For the particles, the unburned fraction is taken as the mass fraction of \ce{Fe} in the particle, and for the gas phase, the normalized (with respect to the inlet) \ce{O2} mass fraction is used.
Again, several observations can be made comparing the 1D to the 3D result.
Until $\eta = \SI{2.5}{\milli\meter}$, the unburned fraction profiles of gas and all particle bins look very similar.
The smallest two bins of particles completely combust within this distance, and the distance required to achieve this combustion is very similar.
Moreover, the normalized oxygen mass fraction reduces to roughly \num{0.5}.
However, the largest bin has barely started burning, so no significant differences can be observed.
Beyond $\eta = \SI{2.5}{\milli\meter}$, large differences are present; the gas phase normalized oxygen profile starts to increase again, and the unburned fraction of the largest particles reduces much more quickly.
Similar to the temperature profiles, this is oxygen from the co-flow diffusing into the unburned mixture.

\begin{figure}
  \centering
  \includegraphics{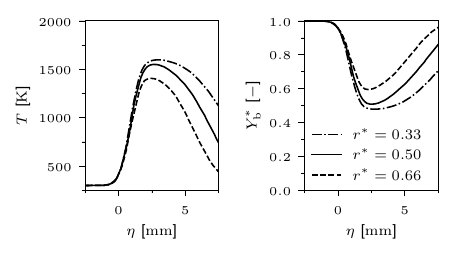}
  \caption{Gas-phase temperature (left) and unburned fraction (right) profiles are shown. 
  Each profile crosses the temperature iso-contour at a different position ($r*=r/R_{\max}$, with $r$ the intersection with the iso-contour) of the reconstructed Bunsen flame.}
  \label{Fig:Radii_flamelets}
\end{figure}

To better demonstrate the impact of the co-flow on the extracted 1D profiles, Fig.~\ref{Fig:Radii_flamelets} is provided.
In the left frame, the gas phase temperature is shown for the flamelet at three radial positions, $r=\num{0.33} R_{\max}$, $\num{0.5} R_{\max}$ and $\num{0.66} R_{\max}$.
As the radius from which the 1D flame is extracted increases, both the peak gas temperature reduces and the cooling rate in the post-flame zone increases.
Below $r = \num{0.5}R_{\max}$, the gas phase temperature profile within the flame front is not altered significantly, while for larger $r$, it is modified.
Similar observations can be made for the unburned fraction profiles in the right frame, except that the minimum of $Y_{\mathrm{b}}^*$ increases for increasing $r$, and the rate at which $Y_{\mathrm{b}}^*$ increases reduces.
Both the temperature profile and the unburned fraction profile align with our earlier observation:
As the radial position of the 1D flame is increased, it is influenced stronger by the cold and oxygen-containing co-flow.
Especially for $r > 0.66 R_{\max}$, a significant portion of the particles combust in an environment that contains more oxygen than the idealized 1D flame, see the left frame of Fig.~\ref{Fig:Unburned_fraction_flamelet} and compare to the right frame of Fig.~\ref{Fig:Radii_flamelets}.
One could expect that this increases the burning velocity, as these flames are known to be sensitive to the oxygen concentration and much less to the gas-temperature due to the diffusion-limited combustion of the particles~\cite{2022_Goroshin_SomeFundamentalAspects}.
Based on this observation, the fact that the burning velocity remains nearly constant between $r/R_{\max}=0.3$ and \num{0.7} in Fig.~\ref{Fig:Flame_speed_1_25} is surprising.
The constant burning velocity suggests that other effects, like varying stretch and curvature, also impact the burning velocity. 

\begin{figure}
  \centering
  \includegraphics{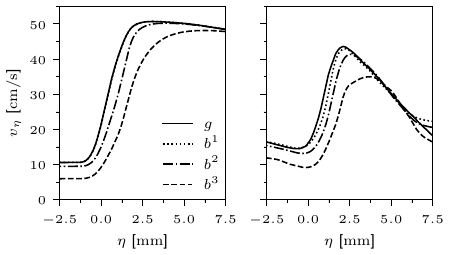}
  \caption{The velocity profile obtained from a 1D flame simulation is shown in the left frame.
  In the right frame, the velocity normal to the temperature iso-contour retrieved from the 3D simulation is shown; see Fig.~\ref{Fig:Reconstructed_flame_front} for the position.
  The particle velocities are shown by binning them: $b^1$ ranges from \num{0.5} to \SI{2.5}{\micro\meter}, $b^2$ ranges from \num{10} to \SI{15}{\micro\meter}, and $b^3$ ranges from \num{20} to \SI{30}{\micro\meter}, and the gas phase velocity is given by the solid line.
  Smoothing using a Gaussian kernel is applied to both gas and particle velocity profiles for both the 1D and 3D results.}
  \label{Fig:velocity_flamelet}
\end{figure}

Finally, profiles of the velocity normal to the flame front are also compared between the 1D and 3D, see Fig.~\ref{Fig:velocity_flamelet}.
When comparing the 1D to the 3D results, several observations can be made.
First, the laminar burning velocity of the 1D flame is lower than that of the 3D flame.
Second, the slip velocity between the various particle bins and the gas phase is similar.
Third, the apparent gas expansion of the gas phase over the flame front is less in the 3D flame ($\pm\num{15}$ to $\pm\SI{45}{\centi\meter\per\second}$, i.e., a factor of roughly three) than in the 1D flame ($\pm\num{10}$ to $\pm\SI{50}{\centi\meter\per\second}$, i.e., a factor of roughly five).
The last observation is interesting, as it can not be explained by the differences of maximum gas phase temperature; see Fig.~\ref{Fig:Temperature_flamelet} and note that the peak temperatures are nearly identical.
As such, the difference in the gas expansion suggests that the extracted 1D flame is significantly impacted by curvature and or stretch.

\subsection{Comparison of burning velocities}
\begin{figure}
  \centering
  \includegraphics{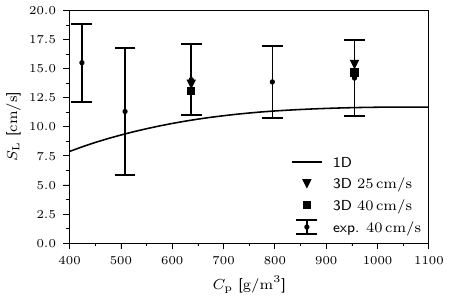}
  \caption{Comparison of the burning velocity from 1D simulations, 3D Bunsen flame simulations at \num{25} and \SI{40}{\centi\meter\per\second}, and experimental results at \SI{40}{\centi\meter\per\second}.
  The burning velocity from the 3D simulations is obtained using the angle method Eq.~\ref{Eq:Flame_speed_angle_based} with the \SI{400}{\kelvin} iso-contour.}
  \label{Fig:Comparison_of_flame_speeds}
\end{figure}

In this final part, the burning velocity estimated from the 3D simulations is compared against the experimental and 1D results, see Fig.~\ref{Fig:Comparison_of_flame_speeds}
For the 3D results, the burning velocity is obtained using the angle method of Eq.~\ref{Eq:Flame_speed_angle_based} and the \SI{400}{\kelvin} iso-contour.
The, until now not presented, simulation results at \SI{40}{\centi\meter\per\second} and $\phi_{\ce{Fe2O3}}=1.0$, and both $\bar{v}=25$ and \SI{40}{\centi\meter\per\second} at $\phi_{\ce{Fe2O3}}=1.5$ are included as well.
Only the experimental results measured at a mean inlet velocity of \SI{40}{\centi\meter\per\second} are included in the graph to prevent the graph from becoming cluttered.
To the authors` knowledge, the recovered burning velocity from the 3D simulation is the best agreement between a model and experiment presented to date for iron dust burning velocities.
It is undoubtedly much better than earlier presented comparisons between 1D models and experiments, often off by more than a factor of two, as observed by Ramaekers et al.~\cite{2025_Ramaekers_influenceradiativeheat}.
Just as observed in the experiment by Fedoryk et al.~\cite{2023_Fedoryk_ExperimentalInvestigationLaminar}, the inlet velocity does not significantly impact the obtained burning velocity.

In contrast to the experiment, we observe a weak dependence of the burning velocity on the particle loading.
However, the variation of the burning velocity as a function of $\phi_{\ce{Fe2O3}}$ is far below the experimental uncertainty, and we can thus not expect it to be measured.
In line with earlier attempts at 1D to experiment comparison, the 1D simulation significantly under-predicts the laminar burning velocity.
We can also provide reasons for the differences based on our analysis of the extracted 1D flames.
\begin{itemize}
  \item Oxygen from the co-flow increases the oxygen concentration in the post-flame zone, accelerating the oxidation of large particles.
  \item Stretch and curvature have some influence on the flame structure. 
\end{itemize}
Only the influence of stretch on the burning velocity has been studied using 1D flames~\cite{2024_Gool_Numericaldeterminationiron}, while the impact of curvature or partial premixing has not yet been studied.
As such, it is, at this stage, unknown what the relative importance of each of these effects is.

\section{Conclusion and Outlook}
In this article, we have performed 3D simulations of the experimental setup presented by Fedoryk et al.~\cite{2023_Fedoryk_ExperimentalInvestigationLaminar}.
With a relatively simple particle model, we have obtained satisfactory agreement between experiment and simulation.
This agreement suggests that the missing factor two, observed in several publications, e.g.,~\cite{2009_Tang_FlamePropagationQuenching,2021_Hazenberg_StructuresBurningVelocities,2025_Ramaekers_influenceradiativeheat}, is due to a poor understanding of the experimental conditions.
Using our results, we have estimated the burning velocity utilizing three different methods:
1) The burning velocity is obtained from the surface area of the flame, 2) the local angle of the flame front is compared against the inlet velocity, and 3) the local gas velocity normal to a temperature iso-contour.
The third method is the only one that can account for the impact of the flame front on the flow field just in front of the flame; as such, this method is expected to be the most reliable.
When these three methods are compared, two conclusions can be made: 1) all three methods have some sensitivity to the utilized temperature iso-contour, but the third method is very sensitive, and 2) the second method produces the highest burning velocity.
When comparing against the experimental results, it is observed that the burning velocity obtained by the third method is very close to the measurement values, providing the best-to-date validation for iron-dust flames.
Still, questions remain about the actual quality of the agreement, as the uncertainty in the experiments is significant, and the impact of various particle models on the flame structure has not been studied yet.
For example, what is the sensitivity to the particle size distribution?
How does the particle seeding quality influence the measured burning velocity?
Do more advanced particle models, i.e., improved ignition~\cite{2022_Mi_quantitativeanalysisignition,2023_Mich_ComparisonMechanisticModels}, oxidation beyond \ce{FeO}~\cite{2023_Gool_ParticleEquilibriumComposition,2024_Mich_Modelingoxidationiron}, diffusion in the free-molecular regime~\cite{2023_Thijs_surfacechemisorptionoxidizing,2023_JeanPhilyppe_ignitionfineiron}, or particle-to-particle radiation~\cite{2025_Ramaekers_influenceradiativeheat}, impact the quality of the agreement?
Or which iso-contour should be taken to most fairly compare against the experiments?
Nevertheless, the agreement obtained in this work provides confidence that the developed models are not missing any significant physics.

Flame structures are compared to better understand the differences between 1D and 3D burning velocities.
From this, we conclude that mixing between the burned gas and the co-flow occurs.
This mixing is large enough that some oxygen from the co-flow can diffuse into the reaction layer of the Bunsen flame, very likely enhancing the flame speed.
Of course, the cold gas from the co-flow also reduces the flame temperature, but iron dust flames are not very sensitive to the flame temperature, e.g.,~\cite{2021_Hazenberg_StructuresBurningVelocities}.
Moreover, we also noted that curvature and stretch effects will likely modify the Bunsen flame`s laminar burning velocity.
It is imperative that the impact of these effects is better understood so that more reliable measurements of the burning velocity can be performed.
Further research of (quasi) 1D flames can be used to improve our understanding.
For example, using 1D counterflow simulations, e.g.,~\cite{2024_Gool_Numericaldeterminationiron}, the effect of stretch and partial-premixing of oxygen can be isolated.
While curvature has been studied with boundary layer resolved simulations by Vance et al.~\cite{2024_Vance_numericalanalysismulti}, no model has been developed to study the impact of curvature on the burning velocity in a 1D framework, such a model can prove crucial to help study the influence of curvature on the burning velocity.

\section{Acknowledgments}
This work was funded by the Hessian Ministry of Higher Education, Research, Science and the Arts - cluster project Clean Circles.

\section{CReDiT authorship contribution statement}
\textbf{T. Hazenberg:} Conceptualization, Data curation, Formal analysis, Investigation, Methodology, Software, Validation, Visualization, Writing - original draft.
\textbf{D. Braig:} Conceptualization, Data curation, Formal analysis, Investigation, Methodology, Software, Validation, Writing - reviewing \& editing.
\textbf{M. A. Fedoryk:} Data curation, Validation, Writing - reviewing \& editing.
\textbf{J. Michs:} Software, Validation, Writing - reviewing \& editing.
\textbf{F. P. Hagen:} Project administration, Supervision, Writing - reviewing \& editing.
\textbf{S. R. Harth:} Project administration, Supervision, Writing - reviewing \& editing.
\textbf{B. Stelzner:} Project administration, Supervision, Writing - reviewing \& editing.
\textbf{A. Scholtissek:} Conceptualization, Funding acquisition, Project administration, Supervision, Writing - reviewing \& editing.
\textbf{D. Trimis:} Conceptualization, Funding acquisition, Project administration, Supervision, Writing - reviewing \& editing.
\textbf{C. Hasse:} Conceptualization, Funding acquisition, Project administration, Resources, Supervision, Writing - reviewing \& editing.

\bibliographystyle{elsarticle-num}
\bibliography{references}

\appendix

\section{Particle temperature equation}
The particle temperature equation can be derived from the enthalpy equation utilized in previous articles, e.g.,~\cite{2021_Hazenberg_StructuresBurningVelocities,2023_Mich_ComparisonMechanisticModels,2023_Gool_ParticleEquilibriumComposition,2024_Gool_Numericaldeterminationiron,2023_Ravi_Effectparticlesize,2023_Ravi_Flamestructureburning}:
\begin{equation}
    \frac{\mathrm{d} H}{\mathrm{d} t} = \dot{q}_{\mathrm{conv}} + \dot{q}_{\mathrm{rad}} + \dot{m}_{\ce{O2}} h_{\ce{O2},\mathrm{s}} \label{Eq:Enthalpy_equation} \mathrm{.}
\end{equation}
The enthalpy can be split into the sensible, $H_{\mathrm{sens}}$, and chemical, $H_{\mathrm{chem}}$, parts,
\begin{equation}
    \frac{\mathrm{d} H_{\mathrm{sens}}}{\mathrm{d} t} + \frac{\mathrm{d} H_{\mathrm{chem}}}{\mathrm{d} t} = \dot{q}_{\mathrm{conv}} + \dot{q}_{\mathrm{rad}} + \dot{m}_{\ce{O2}} h_{\ce{O2},\mathrm{s}} \mathrm{.}
\end{equation}
For a multi-component particle, this can be expanded further into
\begin{align}
        \frac{\mathrm{d} \sum Y_{i} h_{\mathrm{sens},i} m_{\mathrm{p}}}{\mathrm{d} t} 
    & + \frac{\mathrm{d} \sum Y_{i} h_{\mathrm{chem},i} m_{\mathrm{p}}}{\mathrm{d} t} \\
    & = \dot{q}_{\mathrm{conv}} + \dot{q}_{\mathrm{rad}} 
      + \dot{m}_{\ce{O2}} h_{\ce{O2},\mathrm{s}} \mathrm{,} \nonumber
\end{align}
where $Y_i$ is the mass fraction of species $i$ in the particle and $h$ indicates the mass-specific enthalpy.
The first term can be rewritten in terms of temperature by utilizing the chain rule, multiplying with $\mathrm{d} h_{\mathrm{sens}}/\mathrm{d}t$ by $\mathrm{d}T/\mathrm{d}T$, and realizing that $c_p = \mathrm{d} h_{\mathrm{sens}}/\mathrm{d} T$, i.e.,
\begin{align}
       \frac{\mathrm{d} \sum Y_{i} h_{\mathrm{sens},i} m_{\mathrm{p}}}{\mathrm{d} t} 
    &= m_{\mathrm{p}} \sum Y_i c_{\mathrm{p},i} \frac{\mathrm{d} T}{\mathrm{d} t} \label{Eq:sensible_part} \\
    &+ \sum h_{\mathrm{sens},i} \frac{\mathrm{d} m_{\mathrm{p}} Y_i}{\mathrm{d} t}  \mathrm{.} \nonumber
\end{align}
By applying the chain rule, the second term can also be rewritten as
\begin{align}
      \frac{\mathrm{d} \sum Y_{i} h_{\mathrm{chem},i} m_{\mathrm{p}}}{\mathrm{d} t} 
    & = m_{\mathrm{p}} \sum Y_{i} \frac{\mathrm{d} h_{\mathrm{chem},i}}{\mathrm{d} t} \label{Eq:chemical_part} \\
    & + \sum h_{\mathrm{chem},i} \frac{\mathrm{d} m_{\mathrm{p}} Y_{i}}{\mathrm{d} t} \mathrm{,} \nonumber
\end{align}
by definition $\mathrm{d} h_{\mathrm{chem}}/\mathrm{d} t = 0$, such that only the second term remains.
The last term of Eq.~\ref{Eq:sensible_part} and Eq.~\ref{Eq:chemical_part} can be combined to provide the heat-release of reaction:
\begin{align}
     & \sum h_{\mathrm{sens},i} \frac{\mathrm{d} m_{\mathrm{p}} Y_i}{\mathrm{d} t} + \sum h_{\mathrm{chem},i} \frac{\mathrm{d} m_{\mathrm{p}} Y_{i}}{\mathrm{d} t} \label{Eq:heat_release_of_reaction} \\
    =& \sum h_{i} \frac{\mathrm{d} m_{\mathrm{p}} Y_i}{\mathrm{d} t} \nonumber \\
    =& -\dot{q}_{\mathrm{reac}} \mathrm{.} \nonumber
\end{align}
Inserting the above back into the original enthalpy equation (Eq.~\ref{Eq:Enthalpy_equation}) provides
\begin{align}
      \sum Y_i c_{\mathrm{p},i} \frac{\mathrm{d} T}{\mathrm{d} t} 
  & = \dot{q}_{\mathrm{conv}} 
    + \dot{q}_{\mathrm{rad}} 
    + \dot{q}_{\mathrm{reac}} \\
  & + \dot{m}_{\ce{O2}} h_{\ce{O2},\mathrm{s}} \mathrm{,} \nonumber
\end{align}
which is the temperature equation provided in the main text.

\section{Properties of the particle size distribution}
The main article uses the average of three measured particle size distributions.
Here, we will briefly discuss what characteristic diameter is measured for each and provide the fitted distribution's parameters.
We will also provide a small quantitative assessment of the impact of our decision using 1D simulations.

A Camsizer is a dispersion-based particle size and shape characterization device; it works by dispersing a powder sample in the air and taking photographs of the falling powder.
Individual particles are tracked using the photos of the particles, and their size and shape are determined over several frames.
When particles are perfect spheres, their diameter is unique and can easily be determined.
However, the utilized (and most) iron powder is not perfectly spherical; several definitions for the diameter can be used.
For this reason, the sample of the iron powder is measured three times.
Each measurement uses a different diameter definition to determine the particle size distribution.
The three diameter definitions that have been used are:
\begin{enumerate}
    \item $d_{\mathrm{c} \min}$, the particle width.
    The diameter of the particle is determined as the maximum projected chord of all photographs taken from a single particle.
    In the manual of the Camsizer, this definition of the particle size is stated to be close to that obtained from sieving-based methods.
    \item $d_{\mathrm{area}}$, the projected area diameter, computed from
    \begin{equation*}
        d_{\mathrm{area}} = \sqrt{\frac{4 A_{\mathrm{p}}}{\pi}} \mathrm{,}
    \end{equation*}
    where $d_{\mathrm{p}}$ is the projected area of the particle.
    The Camsizer does not specify how an average $d_{\mathrm{area}}$ is determined out of multiple photographs of a single particle. 
    \item $d_{\mathrm{Fe} \max}$, the particle length.
    The diameter is determined as the maximum Feret diameter, the maximum length of the particle, from all the photographs.
    In the Camsizer manual, this definition of particle size is stated to be close to that obtained by microscopy measurements.
\end{enumerate}

\begin{figure*}
    \centering
    \includegraphics{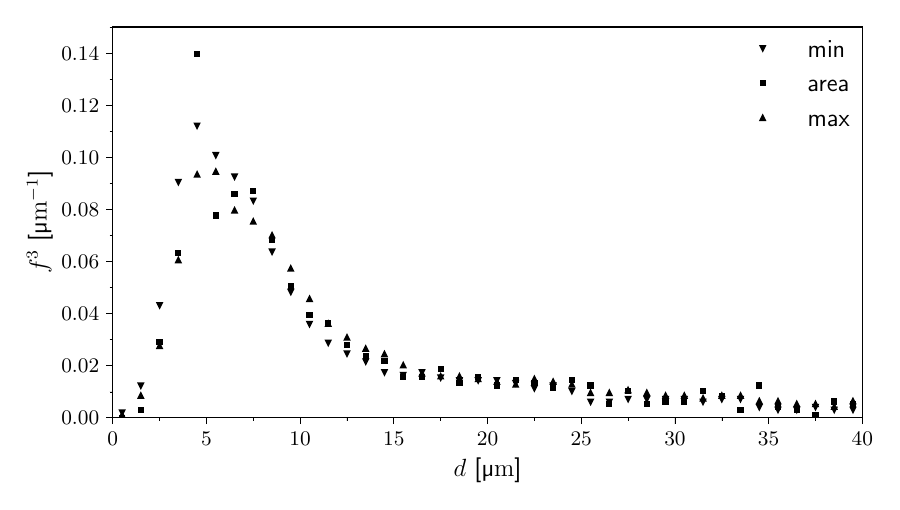}
    \caption{A comparison of the three measured particle size distributions is shown.
    The particle size distribution density function is based on the volume of particles per bin.}
    \label{Fig:Compare_particle_sizes}
\end{figure*}
In Fig.~\ref{Fig:Compare_particle_sizes}, a comparison is made between the three particle size distributions measured with the above three particle size definitions.
Overall, the particle size distributions are comparable, but differences are present.
The particle size distribution in which $d_{\mathrm{c} \min}$ is used for the particle size definition, shows the smallest particles.
Meanwhile, the particle sizes are largest when $d_{\mathrm{Fe} \max}$ is used for the particle size definition.
Even though the used iron powder is close to spherical, the differences are relatively significant and will impact the burning velocity in a numerical simulation.
The Sauter mean diameter of for the $d_{\mathrm{c} \min}$ based distribution is \SI{6.4}{\micro\meter}, the $d_{\mathrm{area}}$ measurement is \SI{7.2}{\micro\meter}, and that of $d_{\mathrm{Fe} \max}$ is \SI{7.4}{\micro\meter}.

The goal of the main article is not to investigate which particle size definition provides the most accurate numerical results.
As such, we have not performed 3D simulations for each particle size definition.
Instead, we fit a log-hyperbolic distribution to the average of the three measured particle size distributions and use this in the 3D simulations.
The average of the three distributions is taken because we did not want to be biased toward one of the three; beyond that, we do not have any physical reason for this choice.
The log-hyperbolic distribution function is provided by
\begin{equation}
    f^n = \mathcal{N} \exp\left( 0.5 C \right) \mathrm{,} \label{Eq:log_hyperbolic_distribution}
\end{equation}
with
\begin{align}
    C & = -\left( \theta + \gamma \right)\sqrt{\delta^2 + (\log\left(d\right) - \mu)^2} \\
      & + \left(\theta - \gamma\right)\left(\log\left(d\right) -\mu\right) \nonumber \mathrm{,}
\end{align}
where $d$ is the particle diameter, $\theta$, $\gamma$, $\mu$, and $\delta$ are distribution parameters, and $\mathcal{N}$ is the normalization constant for the distribution function.
Here, $f^n$ is the probability density distribution function, and $^n$ indicates that this distribution can be fit to number ($^0$), length ($^1$), surface ($^2$) or volume ($^3$) based density function.
In this distribution function $\mu$, controls the average particle size, $\theta$ is the asymptote of the derivative of the $\log(d)$ vs. $\log(p)$ function as $d\to0$, $-\gamma$ is the asymptote for the same function as $d\to\infty$ and $\delta$ controls the width of the distribution.

The distribution function of Eq.~\ref{Eq:log_hyperbolic_distribution} is challenging to fit to the measured particle size distribution.
The best-fit quality is obtained by converting the volume-based density function to a surface-based density function, i.e.,
\begin{equation}
    f^2 = \left(\int \frac{f^3(d)}{d} \mathrm{d} d\right)^{-1}\frac{f^3(d)}{d} \mathrm{.}
    \label{Eq:Volume_to_surface}
\end{equation}
The fit parameters of the distribution Eq.~\ref{Eq:log_hyperbolic_distribution} to the three measured particle size distributions and the average of the three are provided in Table~\ref{Tab:Fit_parameters}.
The parameters are given such that the unit of the resulting function is per meter, and it is the surface density size distribution.
The surface density function can be converted to the volume density function via the inverse of Eq.~\ref{Eq:Volume_to_surface} or to the number density function via
\begin{equation}
    f^0 = \left(\int \frac{f^2(d)}{d^2} \mathrm{d} d\right)^{-1}\frac{f^2(d)}{d^2} \mathrm{.}
\end{equation}
Finally, to provide a sense of the uncertainty due to the particle size distribution, we have computed the laminar burning velocity in a 1D code with particle size distribution.
The computed laminar burning velocity is with gravity ($g=\SI{9.81}{\meter\per\second\squared}$) and without stretch and curvature effects at $\phi_{\ce{Fe2O3}}=1.0$.
For $d_{\mathrm{c} \min}$, $S_{\mathrm{L}} = \SI{11.4}{\centi\meter\per\second}$; for $d_{\mathrm{area}}$, $S_{\mathrm{L}} = \SI{10.7}{\centi\meter\per\second}$; for $d_{\mathrm{Fe} \max}$, $S_{\mathrm{L}} = \SI{10.3}{\centi\meter\per\second}$; and for the average of the three, $S_{\mathrm{L}} = \SI{10.8}{\centi\meter\per\second}$.
Therefore, even if the measurements perfectly characterize the particles, the uncertainty of the numerical burning velocity is around \num{10}\%, which is significant.
\begin{table*}
\centering
\caption{Parameters of fitted particle size distribution functions. 
Note 1) These parameters are based on the surface density size distribution and 2) the unit of the distribution is in per meter and 3) the average is the averaged particle size distribution not the average of the parameters.}
\label{Tab:Fit_parameters}
\begin{tabular}{r|cccc}
                           & $\theta$  & $\gamma$ & $\mu$ & $\delta$ \\ 
    \hline
    $d_{\mathrm{c} \min}$  & \num{1.17} & \num{2.93} & \num{-1.23e1} & \num{2.81e-1} \\
    $d_{\mathrm{area}}$    & \num{3.80} & \num{2.78} & \num{-1.24e1} & \num{3.09e-1} \\
    $d_{\mathrm{Fe} \max}$ & \num{1.26} & \num{2.65} & \num{-1.22e1} & \num{2.25e-1} \\
    average                & \num{1.49} & \num{2.68} & \num{-1.23e1} & \num{1.88e-1} \\       
\end{tabular}
\end{table*}

\end{document}